\documentclass[11pt,twoside]{article}

\usepackage{asp2006}
\usepackage{epsfig}
\usepackage{lscape}
\usepackage{amsmath,amssymb}
\usepackage{natbib}

\newcommand{\tens}[1]{\ensuremath{\boldsymbol{\mathsf{#1}}}}
\newcommand{\vect}[1]{\ensuremath{\boldsymbol{#1}}}
\newcommand{\dd}{\ensuremath{\mathrm{d}}}

\begin{document}
\title{Large Eddy Simulations of Supersonic Turbulence}  
\author{W. Schmidt}  
\affil{Institut f\"ur Astrophysik, Universit\"at G\"ottingen, Friedrich-Hund-Platz 1,
D-37077 G\"ottingen, Germany}    

\begin{abstract} 
Based on ideas by Woodward et al., a subgrid scale model that is applicable to highly compressible turbulence
is presented. Applying the subgrid scale model in large eddy simulations of forced supersonic turbulence, the bottleneck effect is largely reduced and, thereby, approximate scaling laws can be obtained at relatively low numerical resolution. In agreement with previous results from PPM simulations without explicit subgrid scale model, it is found that the energy spectrum function for the velocity field with fractional density-weighing, $\rho^{1/3}\vect{u}$, varies substantially with the forcing, at least for the decade of wavenumbers next to the energy-containing range. Consequently, if universal scaling of compressible turbulence exists, it can be found
on length scales much smaller than the forcing scale only.
\end{abstract}

\section{Introduction}

Large eddy simulations (LES) are of great utility to engineers and atmospheric scientists, but not
commonly used in computational astrophysics. While terrestrial turbulence is incompressible or weakly compressible in most cases, astrophysicists often deal with highly compressible turbulence. Since the stability of numerical solvers for the equations of compressible gas dynamics requires energy dissipation that is intrinsic to the numerical scheme, gas-dynamical simulations at high Reynolds numbers are considered as \emph{implicit} large eddy simulations. However, it was shown that numerical schemes such as the piecewise parabolic method (PPM) of \citet{ColWood84} entail undesired properties of the numerical solutions such as the bottleneck effect, i.~e., an unphysical enhancement of spectral power in the high-wave number range \citep[e.~g.,][]{KritNor07}. This led \citet{WoodPort06} to the idea to couple an \emph{explicit} subgrid-scale (SGS) model to PPM, and they applied the method to decaying transonic turbulence.

In this article, I briefly outline an SGS model that is valid in the supersonic regime. As a first application, I present results from LES of forced supersonic isothermal turbulence with root mean square Mach number between $5$ and $6$, which can be interpreted as an idealized model for cold turbulent gas in the interstellar medium \citep[see][]{KritNor07,FederDuv09}. Computing turbulence energy spectra from these LES, I find that the SGS model largely reduces the bottleneck effect in comparison to plain PPM simulations. The results shed light on the question of the universality of compressible turbulence, which is important for the theory of star formation in molecular clouds \citep[see, for example,][]{ElmeSca04}.

\section{A Subgrid Scale Model for Supersonic Turbulence}

\citet{SchmNie06a} formulated an SGS model based on the energy contained in numerically unresolved
velocity fluctuations, $K_{\mathrm{sgs}}$, to compute turbulent mixing properties. For example, this model improved the treatment of turbulent combustion in thermonuclear supernova simulations. The major modeling challenge is to compute a stress tensor, $\tens{\tau}_{\mathrm{sgs}}$, that accounts for the interaction between numerically resolved scales and the subgrid scales below the grid resolution. The divergence of this stress tensor, $\vect{\nabla}\cdot\tens{\tau}_{\mathrm{sgs}}$, enters as an additional force term in the Euler equation for the conservation of momentum. While the diagonal part of $\tens{\tau}_{\mathrm{sgs}}$ introduces quasi-viscous effects analogous to the viscous term in the Navier-Stokes equation, the diagonal part corresponds to an SGS turbulence pressure that alters the effective equation of state in the highly compressible case.

\citet{WoodPort06} proposed a non-linear closure for $\tens{\tau}_{\mathrm{sgs}}$, which is second order in the trace-free part of the velocity derivative $\vect{\nabla}\otimes\vect{u}$ rather than first order in the trace-free rate of strain (as in the standard eddy-viscosity closure). I investigated a
closure that is a combination of the linear eddy-viscosity closure used in \citet{SchmNie06a} and a non-linear term that is constructed from the complete velocity derivative: 
\begin{equation}
  \label{eq:tau_nonlin}
  \tau_{ij}=C_{1}\Delta\rho^{1/2}K_{\mathrm{sgs}}^{1/2}S_{ij}^{\ast}
  -2C_{2}K_{\mathrm{sgs}}\frac{2u_{i,k}u_{j,k}}{|\vect{\nabla}\otimes\vect{u}|^{2}}
  -\frac{2}{3}(1-C_{2})K_{\mathrm{sgs}}\delta_{ij},
\end{equation}
where $\Delta$ is the size of the grid cells, $\rho$ is the mass density,  $|\vect{\nabla}\otimes\vect{u}|:=(2u_{i,k}u_{i,k})^{1/2}$, and $S_{ij}^{\ast}$ is trace-free symmetric
part of $u_{i,j}$. The SGS turbulence energy density $K_{\mathrm{sgs}}:=-\frac{1}{2}\tau_{ii}$ is given by a dynamical equation that complements the Euler equations for compressible
gas dynamics \citep[see][]{SchmNie06a}. 

In order to verify the applicability of this closure to isotropic supersonic turbulence, \citet{SchmFeder09b} filtered numerical data from $1024^{3}$ PPM simulations \citep{FederDuv09} on a length scale within the approximate inertial subrange. Comparing the explicit computation of the rate of energy transfer, $\Sigma=\tau_{ij}u_{ij}$, from length scales greater than the filter length to smaller length scales to the predictions following from various closures for $\tau_{ij}$, we found that the closure~(\ref{eq:tau_nonlin}) yields the best correlation. The closure
validation as well as further details of the SGS model will be described in a separate article \citep{SchmFeder09b}. 

\begin{figure}[!ht]
\plotone{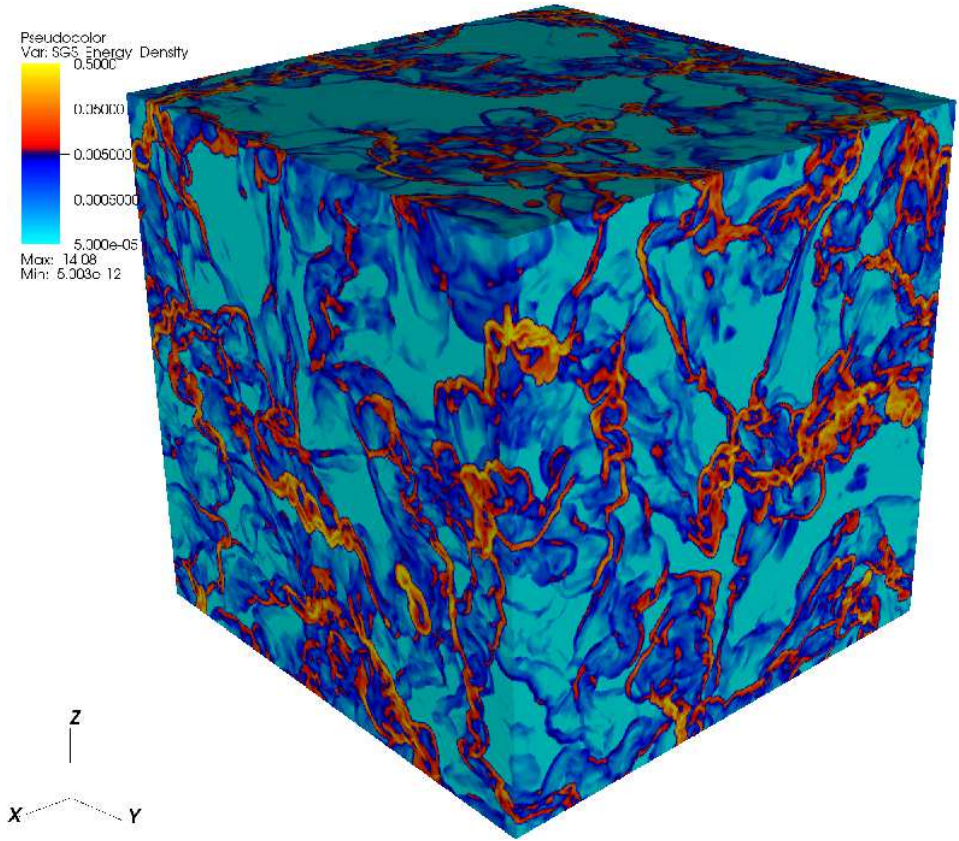}
\caption{\label{fg:ksgs} Visualization of the SGS turbulence energy density $K_{\mathrm{sgs}}$ in a $512^{3}$ LES with purely solenoidal forcing.}
\end{figure}

As an illustration, Fig.~\ref{fg:ksgs} shows a visualization of $K_{\mathrm{sgs}}$ prepared from an LES with $512^{3}$ grid cells. In this simulation, purely solenoidal forcing was applied \citep[the forcing algorithm is described in][]{SchmFeder09a}. In the reddish regions, $K_{\mathrm{sgs}}$ is higher than the average value, while it is lower in bluish regions. It turns out that $K_{\mathrm{sgs}}$ is fairly well correlated with the local denstrophy $\frac{1}{2}\left|\vect{\nabla}\times\left(\rho^{1/2}\vect{u}\right)\right|$, an indicator of compressible turbulent velocity fluctuations \citep{KritNor07}. This reflects the locality of turbulence energy transport from resolved to unresolved scales, although the distribution of $K_{\mathrm{sgs}}$ is smoother in comparison to the local denstrophy, because of the diffusive properties of the SGS dynamics. 

\section{Turbulence Energy Spectra}

It is not obvious which quantity should be considered to specify spectral properties of highly compressible turbulence. Numerical simulations of supersonic turbulence driven by a stirring force showed that the spectrum functions $E(k)$ of kinetic energy per unit mass are significantly stiffer than a Kolmogorov spectrum with slope $-5/3$ \citep[e.~g.,][]{KritNor07}, and indications for a dependence on the forcing were found \citep{SchmFeder09a,FederDuv09}. However, because of the huge fluctuations of the mass density, it seems natural to apply some kind of density weighing. A generalized turbulence energy spectrum function with fractional density weighing, $E_{q}(k)$, is given by integrals over spherical surfaces of radius $k$ in Fourier space,
\begin{equation}
	\label{eq:spect_funct}
	E_{q}(k) = \oint\frac{1}{2}|(\widehat{\rho^{\,q}\vect{u}})^{2}(\vect{k})|k^{2}\,\dd\Omega_{\vect{k}},
\end{equation}
where $\widehat{\rho^{\,q}\vect{u}}$ is the Fourier transform of $\rho^{\,q}\vect{u}$. 
Since the kinetic energy density is given by $\frac{1}{2}\rho u^{2}$, calculating spectrum functions $E_{1/2}(k)$ is an option that suggests itself. Numerical results indicate that the slope of these spectrum functions tend to be flatter than $-5/3$ \citep{KritNor07,FederDuv09}. On the other hand, \citet{Light55} recognized the significance of the rate of dissipation per unit volume for compressible turbulence, which is proportional to the mass density times the third power of the velocity fluctuation on a given length scale. A statistically constant dissipation rate in the inertial subrange implies that the spectrum function $E_{1/3}(k)$ for the density-weighted velocity field $\rho^{1/3}\vect{u}$ has the slope $-5/3$. Following this reasoning, \citet{KritNor07} demonstrated that the spectrum function $E_{1/3}(k)$ computed from a simulation of supersonic isothermal turbulence, indeed, is close to a Kolmogorov spectrum in the inertial subrange. In the following, I present some results concerning the spectral properties of supersonic
turbulence computed by means of LES.

\begin{figure}[!ht]
\plottwo{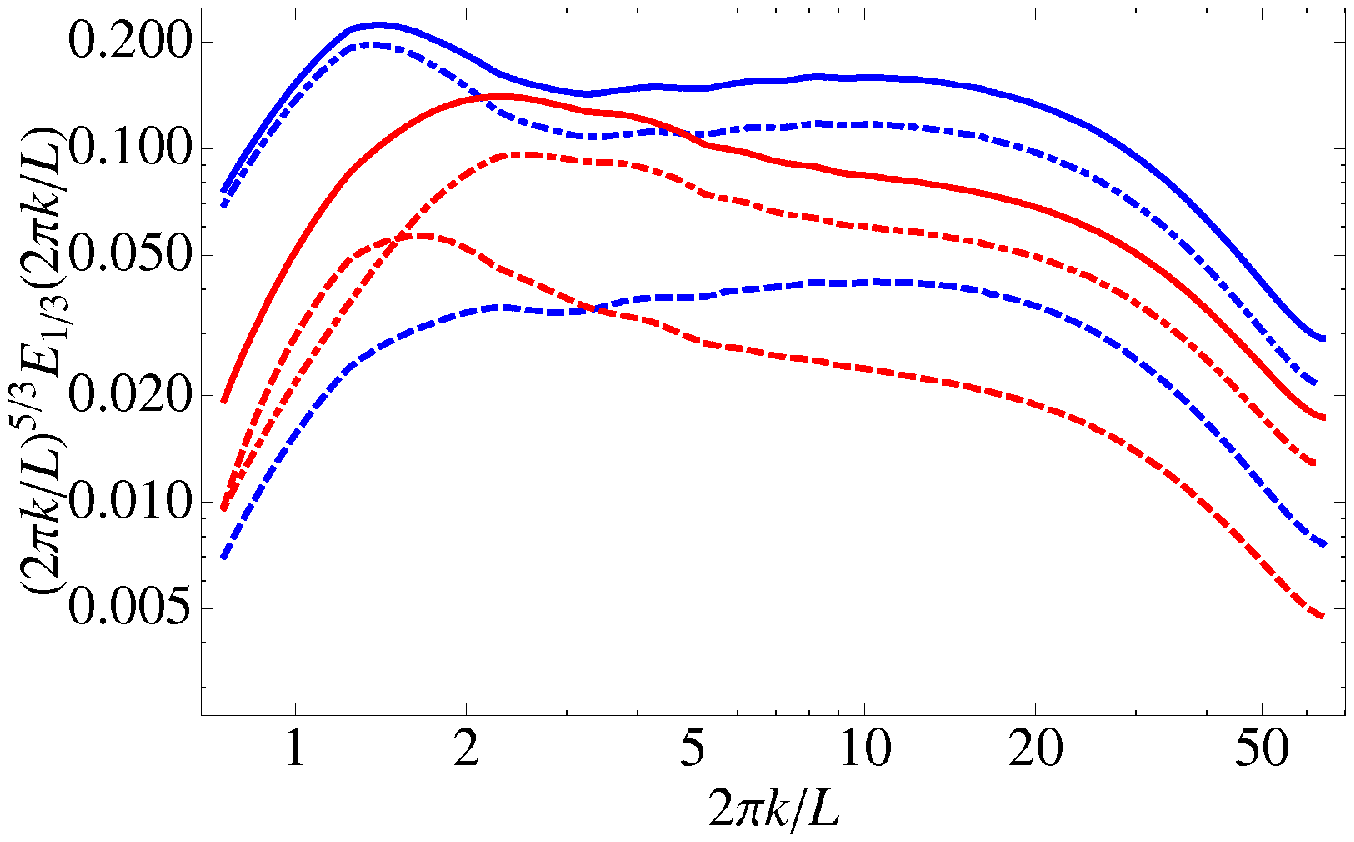}{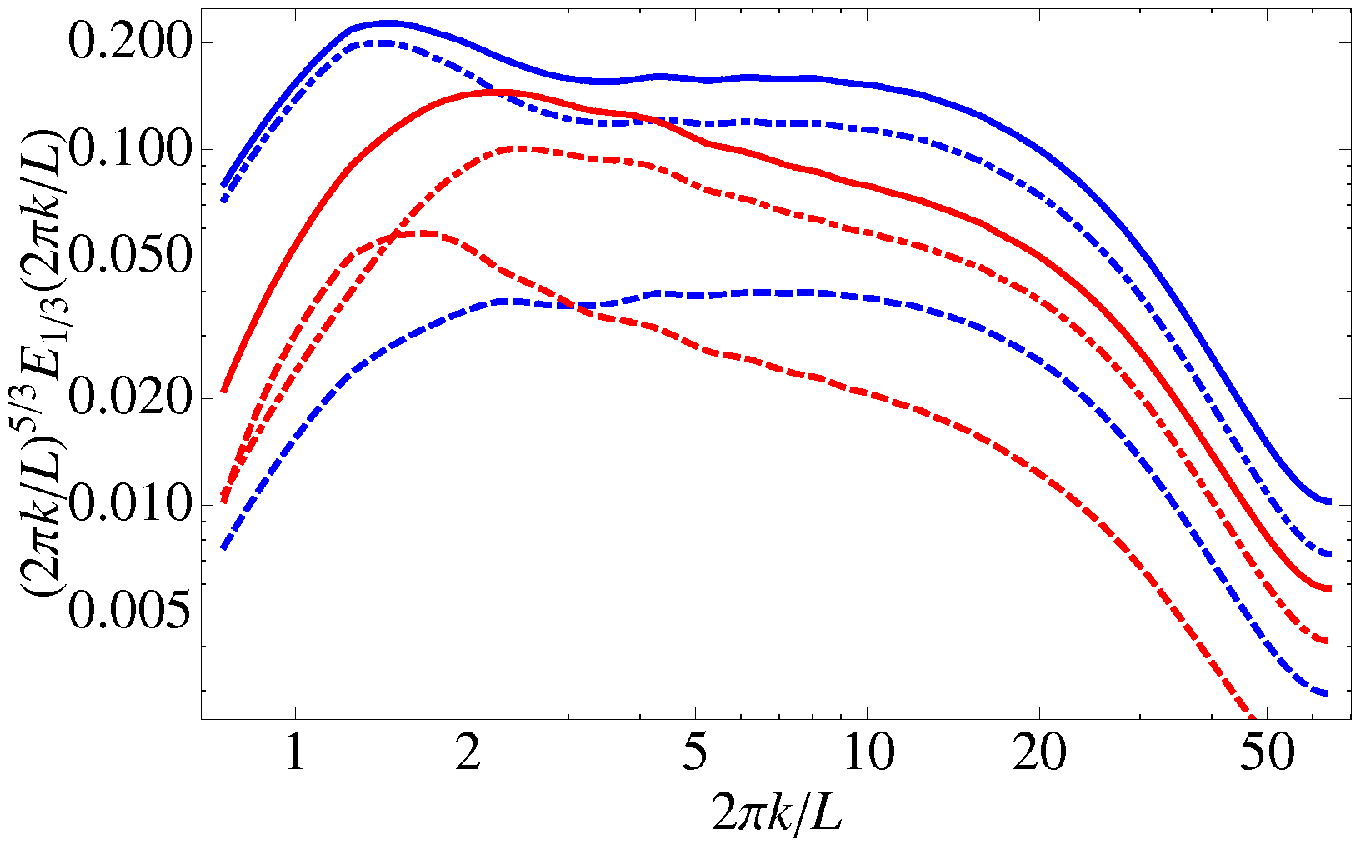}
\caption{Compensated spectrum functions, $k^{5/3}E_{1/3}(k)$, for the mass-weighted velocity field
$\rho^{1/3}\vect{u}$ in plain PPM simulations (left) and LES (right) on $256^{3}$ grid. The spectra for solenoidal and compressive forcing are shown in blue and red, respectively. The dot-dashed lines show transversal components, and the dashed lines show longitudinal components of $E_{1/3}(k)$.\label{fg:spect_256}}
\end{figure}

Let us first compare spectra from LES vs.\ plain PPM simulations on grids with $256^{3}$ cells. As in \citet{FederDuv09}, I consider the two limiting cases of purely solenoidal (divergence-free) and compressive (rotation-free) stochastic force fields. 
Time-averaged compensated spectrum functions, $k^{5/3}E_{1/3}(k)$, calculated from these simulations are plotted in Fig.~\ref{fg:spect_256}. In contrast to the LES spectra (right plot), bumps caused by the bottleneck effect at
high wavenumbers noticeably flatten the spectra obtained from the corresponding PPM simulations without SGS model (left plot). Since the range of wave numbers, for which the spectrum function drops rapidly due to viscous effects, is about the same in both cases, the action of the SGS turbulence stress does not significantly degrade the dynamical range of PPM, which is an important feature. 

\begin{figure}[!ht]
\plottwo{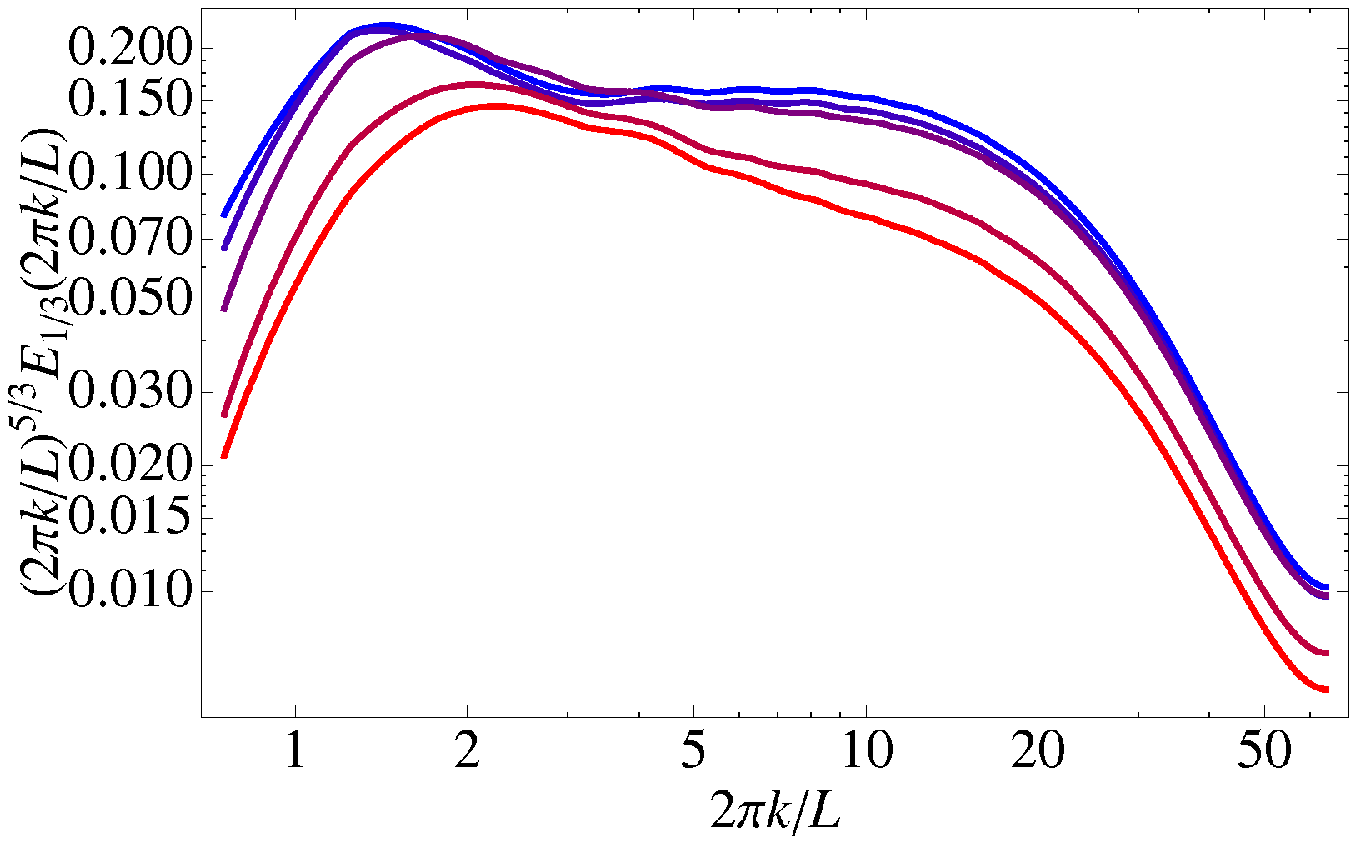}{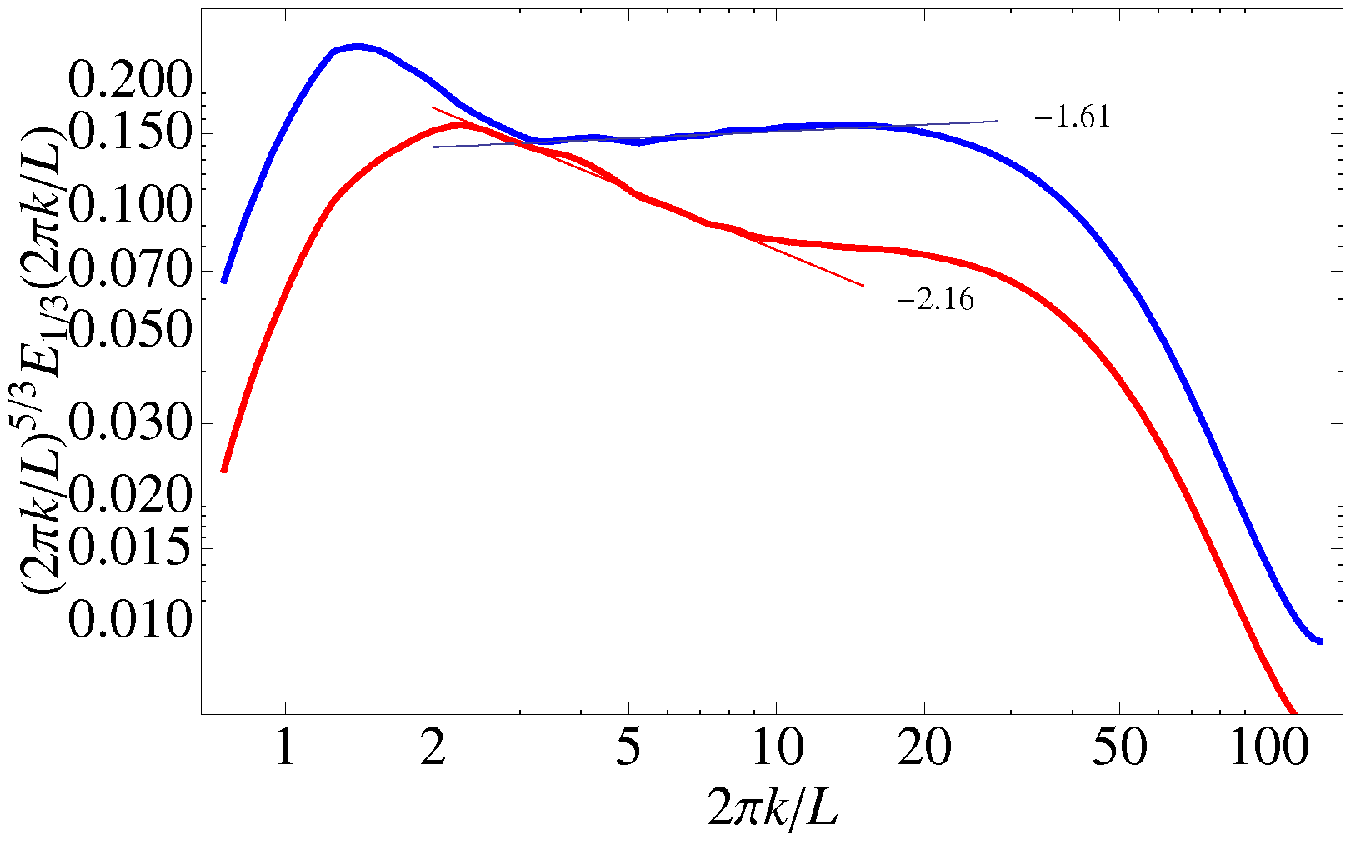}
\caption{Left plot: Results from $256^{3}$ LES, where the weighing parameter $\zeta$
of the forcing \citep[see eqn.~(8) in][]{SchmFeder09a} ranges from $\zeta=1$ (purely solenoidal, shown
in blue) over $2/3$, $1/3$, and $1/6$ to $0$ (purely compressive, shown in red). Right plot: Compensated spectrum functions for $512^{3}$ LES with solenoidal (blue) and compressive (right) forcing.\label{fg:spect_512}}
\end{figure}

Fig.~\ref{fg:spect_256} suggests that in the intermediate wavenumber range $E_{1/3}(k)$ is close to a Kolmogorov spectrum for solenoidal forcing, while it is much stiffer in the case of compressive forcing. One can also see that the contributions from the transversal and the longitudinal components of $\rho^{1/3}\vect{u}$ to the spectrum functions are about the same regardless of the forcing. Spectra for different mixtures of solenoidal and compressive forcing modes are shown in the left plot of Fig.~\ref{fg:spect_512}. The slopes are close to $-5/3$ if the
forcing is mainly solenoidal, but the spectra become substantially steeper if the forcing is dominated by compressive
modes. To determine the slopes of the spectrum functions in between the peaks at low wave numbers and the viscous cutoff at high wavenumbers, I performed LES also on $512^{3}$ grids. Power-law fits $E_{1/3}(k)\propto k^{-\beta_{1/3}}$ yield the spectral indices $\beta_{1/3}=1.61$ and $2.16$ for solenoidal and compressive forcing, respectively (see the right plot in Fig.~\ref{fg:spect_512}). In the latter case, there is a pronounced bend at $2\pi k/L\approx 10$ ($L$ is the forcing length scale), and the spectrum function almost flattens to a slope of $-5/3$ at higher waver numbers. I discuss this feature, which does not emerge until a resolution of at least $512^{3}$ is used, in the concluding Section. In the case of solenoidal forcing, the spectral index slightly decreases at higher resolution, while the shape of the spectrum functions does not change. Nevertheless, since density-weighted velocity scalings tend to be very sensitive on the numerical resolution \citep[see][]{SchmFeder08}, it is encouraging that the spectral indices inferred from the LES agree quite well with the values $\beta_{1/3}=1.64$ (solenoidal forcing) and $2.10$ (compressive forcing) reported by \citet{FederDuv09} for the $1024^{3}$ plain PPM data.

\section{Discussion and Conclusion}

The bottleneck effect, which distorts turbulence energy spectra toward high wavenumbers in PPM simulations without explicit SGS model, is mostly compensated by the SGS turbulence stress in LES. This property is clearly favorable for the determination of scaling laws from the spectra. Moreover, the SGS model for compressible turbulence presented in this article will be valuable in simulations of complex astrophysical flows, where a large range of scales covers energy-injecting physical processes, while only a small portion of the inertial subrange can be resolved. A first example for such an applications has recently been presented by \citet{MaierIap09}

Based on new data for hydrodynamical as well as magnetohydrodynamical turbulence, \citet{KritUst09} put forward the hypothesis that the universality of compressible turbulence becomes manifest in the scalings of $\rho^{1/3}\vect{u}$. The spectral indices inferred from the $512^{3}$ LES presented in this article, however, do not support this hypothesis. There are two interpretations of these results. One interpretation presumes that the scalings in the range of wavenumbers next to the energy-containing range reflect genuine inertial-range properties of supersonic turbulence. Then the turbulence energy spectra clearly exhibit a dependence of these properties on the forcing \citep[also see][]{SchmFeder08}. As proposed by \citet{FederDuv09}, the prominent flattening of the spectrum function $E_{1/3}(k)$ at high wavenumbers that is observed for compressive forcing (see right plot in Fig.~\ref{fg:spect_512}) might be caused by a transition to the weakly compressible regime at the sonic wavenumber, for which the turbulent velocity fluctuations are comparable to the speed of sound. On the other hand, it is possible that compressible turbulence with universal scaling can only develop on length scales much smaller than forcing scale if the forcing is mostly compressive. In order to settle the question which interpretation is correct, the combined influence of varying the Mach number of the flow and choosing different mixing ratios of solenoidal and compressive forcing components has to be investigated. In this regard, the SGS model presented in this article can help to explore the parameter space.

\acknowledgements 

I thank Christoph Federrath for performing the $1024^{3}$ PPM simulations that served as test cases for the LES described in this article and Jens Niemeyer for valuable comments. The SGS model was implemented into the Enzo 1.5 code developed by the Laboratory for Computational Astrophysics at the University of California in San Diego (http://lca.ucsd.edu)

\bibliographystyle{natbib}
\bibliography{Schmidt}

\begin{thebibliography}{}

\bibitem[{Colella} and {Woodward}(1984){Colella} and {Woodward}]{ColWood84}
{Colella}, P. and {Woodward}, P.~R. (1984).
\newblock {The Piecewise Parabolic Method (PPM) for Gas-Dynamical Simulations}.
\newblock {\em Journal of Computational Physics\/}, {\bf 54}, 174--201.

\bibitem[{Elmegreen} and {Scalo}(2004){Elmegreen} and {Scalo}]{ElmeSca04}
{Elmegreen}, B.~G. and {Scalo}, J. (2004).
\newblock {Interstellar Turbulence I: Observations and Processes}.
\newblock {\em \araa\/}, {\bf 42}, 211--273.

\bibitem[{Federrath} {\em et~al.}(2009){Federrath}, {Duval}, {Klessen},
  {Schmidt}, and {Mac Low}]{FederDuv09}
{Federrath}, C., {Duval}, J., {Klessen}, R., {Schmidt}, W., and {Mac Low},
  M.~M. (2009).
\newblock {Comparing the statistics of interstellar turbulence in simulations
  and observations}.
\newblock {E-print arXiv:0905.1060}.

\bibitem[{Kritsuk} {\em et~al.}(2007){Kritsuk}, {Norman}, {Padoan}, and
  {Wagner}]{KritNor07}
{Kritsuk}, A.~G., {Norman}, M.~L., {Padoan}, P., and {Wagner}, R. (2007).
\newblock {The Statistics of Supersonic Isothermal Turbulence}.
\newblock {\em \apj\/}, {\bf 665}, 416--431.

\bibitem[{Kritsuk} {\em et~al.}(2009){Kritsuk}, {Ustyugov}, {Norman}, and
  {Padoan}]{KritUst09}
{Kritsuk}, A.~G., {Ustyugov}, S.~D., {Norman}, M.~L., and {Padoan}, P. (2009).
\newblock {Simulations of Supersonic Turbulence in Molecular Clouds: Evidence
  for a New Universality}.
\newblock In {N.~V.~Pogorelov, E.~Audit, P.~Colella, \& G.~P.~Zank}, editor,
  {\em Astronomical Society of the Pacific Conference Series\/}, volume 406 of
  {\em Astronomical Society of the Pacific Conference Series\/}, pages 15--+.

\bibitem[{Lighthill}(1955){Lighthill}]{Light55}
{Lighthill}, M.~J. (1955).
\newblock {The Effect of Compressibility on Turbulence}.
\newblock In {\em Gas Dynamics of Cosmic Clouds\/}, volume~2 of {\em IAU
  Symposium\/}, pages 121--+.

\bibitem[{Maier} {\em et~al.}(2009){Maier}, {Iapichino}, {Schmidt}, and
  {Niemeyer}]{MaierIap09}
{Maier}, A., {Iapichino}, L., {Schmidt}, W., and {Niemeyer}, J.~C. (2009).
\newblock {Adaptively refined large eddy simulations of clusters}.
\newblock {E-print arXiv:0909.1800}.

\bibitem[{Schmidt} and {Federrath}(2009){Schmidt} and
  {Federrath}]{SchmFeder09b}
{Schmidt}, W. and {Federrath}, C. (2009).
\newblock {In preparation}.

\bibitem[{Schmidt} {\em et~al.}(2006){Schmidt}, {Niemeyer}, and
  {Hillebrandt}]{SchmNie06a}
{Schmidt}, W., {Niemeyer}, J.~C., and {Hillebrandt} (2006).
\newblock {A localised subgrid scale model for fluid dynamical simulations in
  astrophysics. I. Theory and numerical tests}.
\newblock {\em Astron.\ \& Astrophys.}, {\bf 450}, 265--281.

\bibitem[{Schmidt} {\em et~al.}(2008){Schmidt}, {Federrath}, and
  {Klessen}]{SchmFeder08}
{Schmidt}, W., {Federrath}, C., and {Klessen}, R. (2008).
\newblock {Is the Scaling of Supersonic Turbulence Universal?}
\newblock {\em Phys.\ Rev.\ Lett.}, {\bf 101}, 194505.

\bibitem[{Schmidt} {\em et~al.}(2009){Schmidt}, {Federrath}, {Hupp}, {Kern},
  and {Niemeyer}]{SchmFeder09a}
{Schmidt}, W., {Federrath}, C., {Hupp}, M., {Kern}, S., and {Niemeyer}, J.~C.
  (2009).
\newblock {Numerical simulations of compressively driven interstellar
  turbulence. I. Isothermal gas}.
\newblock {\em \aap\/}, {\bf 494}, 127--145.

\bibitem[{Woodward} {\em et~al.}(2006){Woodward}, {Porter}, {Anderson},
  {Fuchs}, and {Herwig}]{WoodPort06}
{Woodward}, P.~R., {Porter}, D.~H., {Anderson}, S., {Fuchs}, T., and {Herwig},
  F. (2006).
\newblock {Large-scale simulations of turbulent stellar convection flows and
  the outlook for petascale computation}.
\newblock {\em Journal of Physics Conference Series\/}, {\bf 46}, 370--384.

\end{thebibliography}



\end{document}